# Perfect Circular Dichroic Metamirrors

Zuojia Wang[1,2,3], Hongsheng Chen[2,3,*], Yongmin Liu[1,4,*]

[1]*Department of Mechanical and Industrial Engineering, Northeastern University, Boston, MA, 02115, USA.*
[2]*Department of Information Science and Electronic Engineering, Zhejiang University, Hangzhou 310027, China.*
[3]*The Electromagnetics Academy at Zhejiang University, Hangzhou 310027, China.*
[4]*Department of Electrical and Computer Engineering, Northeastern University, Boston, MA, 02115, USA.*

**Abstract**

In nature, the beetle *Chrysina gloriosa* derives its iridescence by selectively reflecting left-handed circularly polarized light only. Here, for the first time, we introduce and demonstrate the optical analogue based on an ultrathin metamaterial, which we term circular dichroic metamirror. A general method to design the circular dichroic metasmirror is presented under the framework of Jones calculus. It is analytically shown that the metamirror can be realized by two layers of anisotropic metamaterial structures, in order to satisfy the required simultaneous breakings of *n*-fold rotational ($n>2$) and mirror symmetries. We design an infrared metamirror, which shows perfect reflectance for left-handed circularly polarized light without reversing its handedness, while almost completely absorbs right-handed circularly polarized light. These findings offer new methodology to realize novel chiral optical devices for a variety of applications, including polarimetric imaging, molecular spectroscopy, as well as quantum information processing.

[*]Corresponding author. Email: y.liu@neu.edu, hansomchen@zju.edu.cn

In addition to the intensity, wavelength and phase, the polarization of light plays a critical role in light-matter interactions. Biological species show remarkable examples in this regard. For instance, *Gonodactylus smithii* (mantis shrimps) can unambiguously distinguish left-handed circularly polarized (LCP) and right-handed circularly polarized (RCP) light [1]. *Chrysina gloriosa* (jeweled beetles) under LCP illumination appear more brilliant than those under RCP illumination [2]. In quantum information technology, manipulating the spin state of electrons by using circularly polarized light is also of great importance [3-6]. As a result, the efficient analysis and engineering of the polarization state is imperative in diverse disciplines, including biology, physics, materials science and optical communications. Because of the extraordinary capability of creating novel material properties and controlling light flow in a prescribed manner, metamaterials have intrigued tremendous interests over the past decades, and become a popular strategy to obtain enormous chiral responses, such as negative refractive index [7, 8], giant gyrotropy [9] and strong circular dichroism [10, 11]. The recently developed ultrathin, planar metamaterials, known as metasurfaces, promise simple and low-loss designs to control the polarization [12-14], effectively preserve the handedness in reflection [15] and manipulate the ray trajectory of circularly polarized light [16-18]. Nevertheless, metasurfaces typically requires the gradient phase shift of neighboring artificial structures, and so far they are not able to selectively operate on a desired circular polarization state.

In this Letter, we demonstrate, for the first time, a new concept of circular dichroic (CD) metamirrors, the optical analogue of *Chrysina gloriosa* in nature that allows selective reflection of a designated circularly polarized incident waves without reversing its handedness, while the other polarization state is completely absorbed. To achieve this function, Jones calculus is firstly employed to predict the required reflection matrix. The subsequent analyses in structural symmetry show that we need to break both $n$-fold rotational ($n>2$) and mirror symmetries. Based on the transfer matrix approach, a proof-of-concept CD metamirror utilizing an ultrathin, bi-layer metamaterial is then designed and optimized in the infrared region. Simulation results show 94.7% reflectance and 99.3% absorption for LCP and RCP waves, respectively. In addition, the device performance is insensitive to the incident angle and the offset between the two layers of metamaterial structures. Such extremely high extinction ratio as well as geometry and illumination tolerance manifests promising potential applications in polarimetric imaging, CD spectroscopy, and quantum optical information processing. Last but not least, the planar metamirror is only quarter-wavelength thin, which is highly desirable for on-chip device integration.

The concept of CD metamirrors is illustrated in Fig. 1. For an ordinary metallic mirror, reversal of handedness occurs when a circular polarized light is reflected off its surface at normal incidence. For oblique incidence, the reflected light is elliptically polarized in general. This phenomenon dramatically complicates the optical setup and characterization when circularly polarized light is involved. In contrast, the function of the proposed CD metamirror is to selectively reflect a certain circularly polarized wave and preserve its handedness over a relatively wide range of incident angles. For simplicity, selective reflection of LCP wave is considered in the following. This means normally incident LCP light is totally reflected by the metamirror, while the RCP state is perfectly absorbed. This property is distinctly different from the previous metasurface design, which acts as a half wave plate to convert the handedness of the reflected wave, no matter it is LCP or RCP [15].

The functionality of the CD metamirror is urgently needed in optical engineering. Traditionally, circularly polarized light is generated by the combination a linear polarizer and a quarter-wave plate, which are bulky. Multiple reflections from a series of metallic mirrors are often unavoidable in an optical setup. Consequently, optical signals can hardly preserve their original circular polarization states after many reflection interfaces. The proposed metamirror can be an ideal strategy to address these issues. A circularly polarized beam can be generated once linearly polarized light is reflected from the ultrathin CD metamirror. Furthermore, the preservation of the handedness after reflection also helps to suppress the mode coupling between two circular polarization states and reduce the complexities in optical instruments.

In the following, we first analyze the required condition for selective reflection of one designed circular polarization state based on Jones calculus. Consider two half-spaces separated by a metamirror at $z=0$. The fields in the two regions can be related via Jones calculus [19]

$$\begin{pmatrix} E_r^x \\ E_r^y \end{pmatrix} = \begin{pmatrix} r_{xx} & r_{xy} \\ r_{yx} & r_{yy} \end{pmatrix} \begin{pmatrix} E_i^x \\ E_i^y \end{pmatrix} = \mathbf{R} \begin{pmatrix} E_i^x \\ E_i^y \end{pmatrix}, \quad (1)$$

$$\begin{pmatrix} E_t^x \\ E_t^y \end{pmatrix} = \begin{pmatrix} t_{xx} & t_{xy} \\ t_{yx} & t_{yy} \end{pmatrix} \begin{pmatrix} E_i^x \\ E_i^y \end{pmatrix} = \mathbf{T} \begin{pmatrix} E_i^x \\ E_i^y \end{pmatrix}, \quad (2)$$

where $\mathbf{R}$ and $\mathbf{T}$ are the reflection and transmission matrices of the metamirror, while $E_i^{x,y}$, $E_r^{x,y}$ and $E_t^{x,y}$ are the incident, reflected and transmitted fields, polarized in the $x$ and $y$ directions, respectively. Consequently, RCP and LCP waves can be represented by $(1, i)^T$ and $(1, -i)^T$, respectively, The transmission matrix must equal zero ($\mathbf{T} = 0$) for an ideal, reflecting mirror. For the reflection matrix $\mathbf{R}$, we can readily prove that complete reflection of LCP waves and total absorption of RCP waves require [supplemental material]

$$\mathbf{R} = \begin{pmatrix} r_{xx} & r_{xy} \\ r_{yx} & r_{yy} \end{pmatrix} = \frac{e^{i\alpha}}{2} \begin{pmatrix} 1 & i \\ i & -1 \end{pmatrix}, \quad (3)$$

where $\alpha$ is an arbitrary phase shift through the metamirror and a time-harmonic propagation of $e^{-i\omega t}$ is assumed. To characterize the optical behavior, it is useful to analyze the eigenstates of the polarization, which are uniquely related to the symmetry. By solving a simple eigenvalue problem, we obtain the eigenvalue $\kappa = 0$ with the eigenvector $(1, i)^T$, corresponding to a RCP state as defined from the point of view of the source. This implies that no light is reflected at all if the metamirror is illuminated by RCP light. In contrast, LCP incident waves are completely reflected, and its handeness is preserved.

To better understand the physical meaning of the desired reflection matrix in Eq. (3), now we perform symmetry analysis and identify which kind of structural symmetry should be satisfied for our CD metamirrors. Mathematically, rotating the structure by an arbitrary angle with respect to the z-axis can be accomplished by applying the following matrix operation [19]

$$\mathbf{R}_{new} = \mathbf{D}_\varphi^{-1} \mathbf{R} \mathbf{D}_\varphi, \text{ with } \mathbf{D}_\varphi = \begin{pmatrix} \cos(\varphi) & \sin(\varphi) \\ -\sin(\varphi) & \cos(\varphi) \end{pmatrix}, \quad (4)$$

where $\varphi$ is the rotation angle and $\mathbf{R}_{new}$ is the new reflection matrix of the rotated sample. If the structure has a certain rotational symmetry, the invariance of reflection matrix under transformation requires $\mathbf{R}_{new} = \mathbf{R}$. By substituting Eq. (3) into Eq. (4), the condition of rotational symmetry can be written as

$$\sin(\varphi) \begin{pmatrix} r_{xy} + r_{yx} & r_{yy} - r_{xx} \\ r_{yy} - r_{xx} & -r_{xy} - r_{yx} \end{pmatrix} = \mathbf{0}. \quad (5)$$

To satisfy both the desired reflection matrix in Eq. (3) and the rotational symmetry condition in Eq. (5), the only solution is $\varphi = m\pi, \ m = 0, \pm 1, \cdots$. Hence, to achieve ideal CD metamirrors, only the $C_2$ symmetric group is allowed.

Other than rotational symmetry, another basic characteristic of artificial structures is the mirror symmetry. Similarly, we can also employ the method of matrix transformation to predict the requirement of mirror symmetry for ideal CD metamirrors. If the metamirror is mirror-symmetric with respect to a plane rotated from the *xz* plane by an angle $\varphi$ along the *z*-axis, the reflection matrix for the structure reflected at that plane is identical to the original one. Therefore we have

$$\mathbf{R}_{new} = \mathbf{A}_x^{-1} \mathbf{D}_{-\varphi}^{-1} \mathbf{R} \mathbf{D}_{-\varphi} \mathbf{A}_x, \qquad (6)$$

where $A_x = \begin{pmatrix} 1 & 0 \\ 0 & -1 \end{pmatrix}$ is the mirror matrix with respect to the *x*-axis, $\mathbf{D}_{-\varphi}$ is given by replacing $\varphi$ to $-\varphi$ in $\mathbf{D}_\varphi$. By applying the symmetry condition $\mathbf{R}_{new} = \mathbf{R}$ again, we obtain

$$\sin(2\varphi)(r_{xx} - r_{yy}) + 2\cos(2\varphi) r_{xy} = 0. \qquad (7)$$

According to Eq. (3), the ratio between $r_{xx} - r_{yy}$ and $r_{xy}$ is a pure imaginary number so that the mirror-symmetric condition described by Eq. (7) can never be met. Hence, selective reflection of circular polarization does not exist in any kind of structure that contains mirror symmetry with respect to the plane of incidence.

Aforementioned analyses point out a general rule to design CD metamirrors: simultaneous breakings of *n*-fold rotational (*n*>2) and mirror symmetries. A straightforward strategy to meet the symmetric conditions is combining two anisotropic metamaterial interfaces with a relative twisting angle, as shown in Fig. 2. Here, a continuous metallic wire array, probably the simplest structure for anisotropy, is adopted as the top layer and the bottom layer consists of a metallic rod placed above an optically thick metallic reflector. The total transmission is hence zero. A dielectric material with refractive index $n_d$ is used as the spacer between the top metallic wires and the bottom metallic reflector. We apply the transfer matrix method in the CD metamirror design and optimization. In general, for a stratified planar structure between two media *a* and *b*, the 4×4 matrix **M** relates forward and backward propagating fields [20]

$$\left(E_{bx}^{(f)}, E_{by}^{(f)}, E_{bx}^{(b)}, E_{by}^{(b)}\right)^T = \mathbf{M} \left(E_{ax}^{(f)}, E_{ay}^{(f)}, E_{ax}^{(b)}, E_{ay}^{(b)}\right)^T. \qquad (8)$$

Here the superscripts (*f*) and (*b*) denote the forward and backward propagating light, respectively. The overall transfer matrix **M** can be expressed as a multiplication of the transfer matrix of each layer [supplemental material]. Once the transfer matrix is fixed, the reflection matrix for the entire structure is then given by

$$\mathbf{R} = -\begin{pmatrix} m_{33} & m_{34} \\ m_{43} & m_{44} \end{pmatrix}^{-1} \begin{pmatrix} m_{31} & m_{32} \\ m_{41} & m_{42} \end{pmatrix}, \qquad (9)$$

where $m_{nl}$ is the *n*-th row and *l*-th column component of the overall 4×4 transfer matrix **M**. Therefore, by combining the overall transfer matrix with appropriate optimization algorithm, the desired thickness of the dielectric spacing layer can be determined.

The proof-of-concept circular dichroic metamirror is numerically demonstrated in the infrared frequency region. The width of both the top metallic wire and the bottom metallic rod is $w = 0.2$ $\mu m$ and the length of later is $l = 2~\mu m$, with the period of $p = 2.5~\mu m$. The metallic rod is rotated by $\pi/4$ with respect to z-axis to ensure no rotational and mirror symmetries. A dielectric material with refractive index $n_d = 1.51$ is used to separate different metal layers with the thickness $d = 1.55~\mu m$ and $t = 0.09~\mu m$. Gold is employed as the metal, and the thickness of the metallic structures is $0.1~\mu m$. The permittivity of gold is described by Drude model after fitting the experimental data from the literature [21].

We validate the concept of selective reflection for LCP waves by performing full-wave numerical simulations (CST Microwave Studio). As shown in Fig. 3a, the absorption of the structure is as high as 99.3% for RCP light, while only 5.3% for LCP light at the wavelength of $8.1~\mu m$. Simulation results match extremely well with theoretical prediction based on the transfer function, due to the low coupling between the top and bottom layers. If the thickness of the middle dielectric reduces to deep subwavelength scales, the interactions between top and bottom gold structures gradually dominates and leads to slight discrepancy [supplemental material]. Moreover, if we focus on the reflection spectrum presented in Fig. 3b, it is noticeable that, for the LCP incident case, the magnitude ratio between the reflected LCP and RCP light reaches up to 14.8. Therefore, the proposed metamirror can not only selectively reflect the LCP waves with high efficiency but also preserve the handedness, which is distinctly different from previous metasurface work [15, 18].

The underlying physical mechanism of selective reflection originates from the interference of multiple reflections. As we know, LCP and RCP can be considered as mirror images of each other with respect to the yz-plane. The mirror symmetry existed in the top layer guarantees the identical optical behavior for two circular polarization states. Nevertheless, the lack of yz-plane mirror symmetry in the bottom layer gives different phase shifts for two states, similar to concept of Pancharatnam-Berry phase in metasurfaces [16]. Consequently, the constructive and destructive interferences occur for LCP and RCP lights, respectively.

Another remarkable feature of our CD metamirror is the good performance for a wide range of incident angles. The angular sensitivity of the metamirror is investigated for two different cases for the wave vector confined in the xz-plane and yz-plane, respectively. One can clearly see from Fig. 4 that the operating wavelength does not significantly change under oblique incidences. Moreover, the absorption of two polarization states still preserves high contrast (>79%) for an incident angle range up to 40 degree. The main reason for this feature is due to the angular insensitivity of resonance frequency in the bottom metamaterial structure [supplemental material]. This omnidirectional performance of the CD metamirror is definitely beneficial in its application as a single mode mirror for circular polarization lights. We further investigate the offset influence of the top layer, which may occur in practical fabrications. As illustrated in Fig. 5, the performance of the metamirror is fairly stable when the top wire has a lateral offset up to 0.1 μm with respect to the bottom rod. This feature is understandable, because the current structure works in the weak-coupling regime. The top layer can be always considered as homogeneous anisotropic metasurface no matter how much distance it is offset.

In conclusion, we have proposed a perfect circular dichroic metamirror to mimic the iridescent attributes of the beetles in nature. The breaking of rotational and mirror symmetries is theoretically proven as the necessary condition in the structure design. By employing the transfer matrix approach, a metamirror comprising two twisted anisotropic metamaterial interfaces is designed and optimized in infrared region. It is shown that such metamirror can almost perfectly reflect all the LCP lights with preservation of the handedness and totally absorb the RCP incidences. The prominent performance in angular sensitivity indicates promising applications in characterization of circularly polarization states. Independence on offset between two layers can reduce the difficulty in fabrication. Furthermore, the principle we applied here is universal. We therefore expect this investigation opens a gateway to achieve more single mode devices for circular polarization lights and should be beneficial for the development of photonics.

Zuojia Wang would like to acknowledge the Chinese Scholarship Council (CSC, No. 201406320105) for financial support.

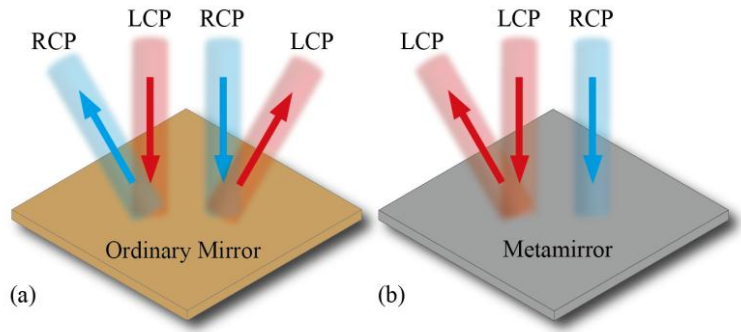

FIG. 1. (a) Ordinary mirrors reverse the handedness of the circularly polarized waves in reflection. (b) CD metamirrors reflect LCP waves only and preserve the handedness.

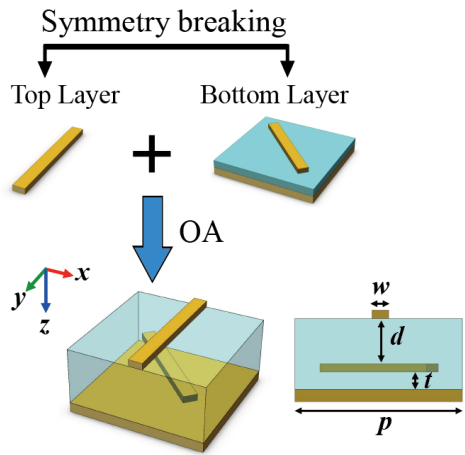

FIG. 2. Schematic of the design procedure for CD metamirrors. Twisted metamaterial interfaces are constructed to break desired structural symmetries. Thickness of the middle substrate is optimized based on transfer matrix approach.

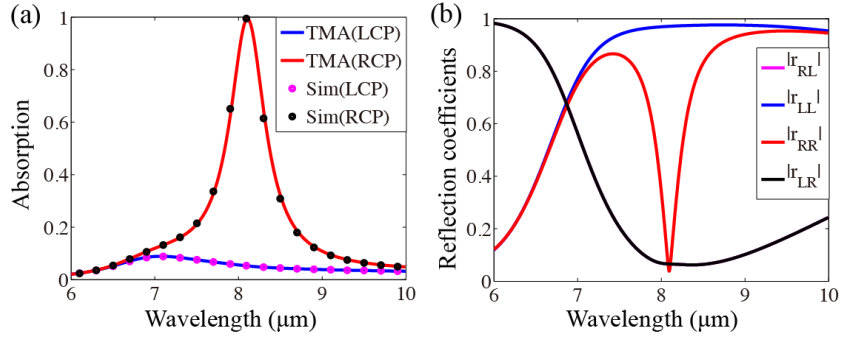

FIG. 3. (a) Simulated and calculated absorption spectra of the CD metamirror. Calculated results are obtained based on transfer matrix approach (TMA). (b) Simulated reflection coefficients of the CD metamirror for circularly polarization incidences.

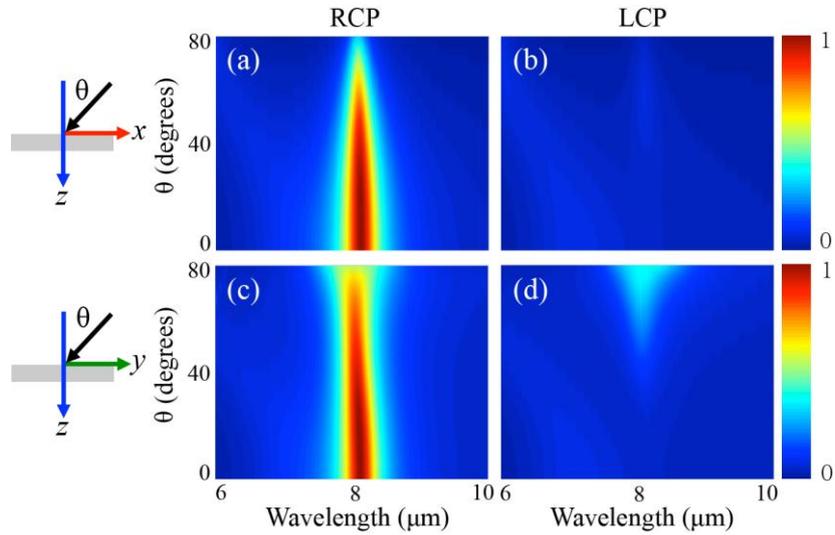

FIG. 4. Simulated absorption of (a, c) RCP and (b, d) LCP states when the incident lights propagate in *xy*-plane or *yz*-plane, respectively.

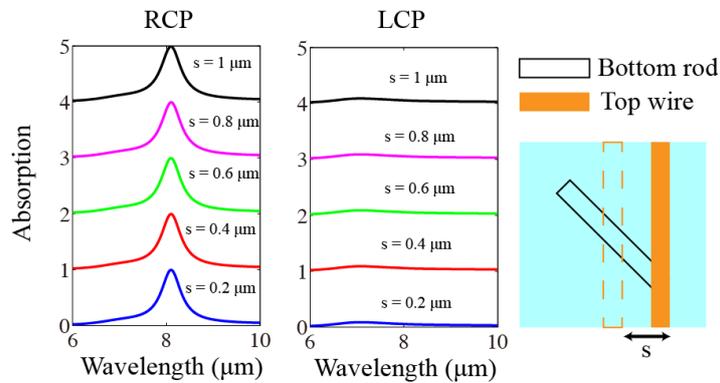

FIG. 5. Dependence of the absorption for (a) RCP and (b) LCP illuminations on the offset distance of the top wire.

- 7 -